\documentclass[preprint,prd,aps,showpacs,preprintnumbers,amsmath,amssymb]{revtex4-1}
\usepackage{graphics,epsfig,subfigure}
\usepackage{diagbox}
\usepackage[usenames]{color}
\usepackage[colorlinks,
            linkcolor=black,
            anchorcolor=blue,
            citecolor=blue
            ]{hyperref}
\usepackage{color}
\usepackage{graphicx}
\usepackage{amsfonts}
\usepackage{indentfirst}
\usepackage{booktabs}

\begin{document}
\renewcommand{\baselinestretch}{1.3}
\newcommand\beq{\begin{equation}}
\newcommand\eeq{\end{equation}}
\newcommand\beqn{\begin{eqnarray}}
\newcommand\eeqn{\end{eqnarray}}
\newcommand\nn{\nonumber}
\newcommand\fc{\frac}
\newcommand\lt{\left}
\newcommand\rt{\right}
\newcommand\pt{\partial}

\title{\Large \bf Chains of Rotating mini-Boson Stars}
\author{Shi-Xian Sun and Yong-Qiang Wang\footnote{yqwang@lzu.edu.cn, corresponding author}}

\affiliation{$^{1}$Institute of Theoretical Physics and Research Center of Gravitation, Lanzhou University, Lanzhou 730000, China\\
    $^{2}$Lanzhou Center for Theoretical Physics and Key Laboratory of Theoretical Physics of Gansu Province, Lanzhou University, Lanzhou 730000, China}

\begin{abstract}
In this article, we investigate the stationary, soliton-like solutions in the model of the Einstein gravity coupled to a free and complex scalar field, and extend chains of mini-boson stars to the rotating case. These solutions manifest as multiple rotating mini-boson stars uniformly arranged along the rotation axis. Through numerical methods, we obtain chains of rotating mini-boson stars with one to five constituents.  We show the distribution of the field functions for these chain solutions. Additionally, we also study the effect of the frequency of the complex scalar field on the ADM mass $M$ and angular momentum $J$. By comparing the conclusions of the rotating case with the non-rotating case, there are some intriguing differences. Furthermore, we observe that there exist two ergospheres for some of these solutions.
\end{abstract}

\maketitle

\section{Introduction}\label{Sec1}

Boson stars, as localized stationary solutions composed of classical fields and Einstein gravity, originated from Wheeler's concept of ``geons" \cite{Wheeler:1955zz}. Geons were energy-localized configurations without singularities formed by electromagnetic or gravitational waves interacting with their own gravity. However, such solutions were not found. Subsequently, Kaup replaced the electromagnetic field with a massive complex scalar field, resulting in the formation of Klein-Gordon geons \cite{Kaup:1968zz} (i.e., boson stars). R. Ruffini and S. Bonazzola independently derived boson star solutions \cite{Ruffini:1969qy}. The stable existence of these solutions depends on the balance between the dispersion of the scalar modes and the gravitational attraction. Boson stars can attain macroscopic sizes and masses, even exhibit high compactness. Over the years, various types of boson star solutions have been obtained, including rotating \cite{Schunck:1996he, Ryan:1996nk, Yoshida:1997qf}, charged \cite{Jetzer:1989av, Jetzer:1989us, Pugliese:2013gsa, Kumar:2017zms, Collodel:2019ohy, Lopez:2023phk, Jaramillo:2023lgk, Brihaye:2023hwg}, excited states \cite{Balakrishna:1997ej, Bernal:2009zy, Collodel:2017biu, Wang:2018xhw, Herdeiro:2020kvf}, self-interacting \cite{Friedberg:1986tq, Lee:1986ts, Colpi:1986ye, Schunck:1999zu, Barranco:2010ib, Grandclement:2014msa, Guerra:2019srj, Li:2020ffy, Delgado:2020udb}, and multifield \cite{Alcubierre:2018ahf, Zeng:2021oez, Sanchis-Gual:2021edp, Dzhunushaliev:2021vwn, Liang:2022mjo, Ma:2023vfa, Pombo:2023sih}, building upon the spherically symmetric ground state mini-boson stars. Additionally, boson stars serve as mimickers of black holes \cite{Torres:2000dw, Vincent:2015xta, Cardoso:2019rvt, Herdeiro:2021lwl, Rosa:2023qcv}, candidates for dark matter \cite{Matos:2007zza, Eby:2015hsq, Sahni:1999qe, Padilla:2019fju}, and sources of gravitational waves \cite{Cardoso:2016oxy, CalderonBustillo:2020fyi, Zhang:2021ojz, Bezares:2022obu}.

Recently, Ref. \cite{Herdeiro:2021mol} re-investigate the soliton solutions in the model of the Einstein gravity coupled to a complex scalar field with the self-interaction potential with the type of quartic and sextic terms. They have discovered a new family of solutions in addition to the typical spherically symmetric boson star solutions. These solutions characterized by multiple boson stars arranged in a continuous sequence with similar sizes and distances, so called ``chains of boson stars". In flat spacetime, scalar fields with the self-interaction potential containing quartic and sextic terms can also have soliton solutions, known as $Q$-balls \cite{Friedberg:1976me, Deppert:1979au, Mielke:1980sa, Coleman:1985ki, Lee:1991ax, Kleihaus:2007vk, Volkov:2002aj, Kleihaus:2005me}, named after the conserved Noether charge ``$Q$" associated with the particle number. $Q$-balls, along with many other self-interacting solitons, allow for chain configurations \cite{Kleihaus:2003nj, Kleihaus:2004is, Kunz:2006ex, Shnir:2007zz, Foster:2009vk, Shnir:2009ct, Ibadov:2010hm, Loiko:2020htk}. The presence of Einstein gravity implies that self-interaction is not a necessary condition for soliton solutions. We has been demonstrated that scalar fields without self-interaction also give rise to solutions with chain configurations \cite{Sun:2022duv}. It is noteworthy that chains of boson stars with two constituents as $2P$ state boson stars \cite{Kleihaus:2007vk, Wang:2018xhw, Kunz:2019bhm, Cunha:2022tvk} (also referred to as dipolar boson stars), have been researched extensively.

Rotation is a ubiquitous phenomenon in the natural world. Ref. \cite{Gervalle:2022fze} extends the solutions of Ref. \cite{Herdeiro:2021mol}, obtaining chains of rotating boson stars with the self-interaction potential with the type of quartic and sextic terms. The primary objective of this paper is to obtain chains of rotating mini-boson stars without self-interaction and analyze their properties. For the non-rotating case, chains of boson stars with an odd number of constituents exhibit different characteristics from those with an even number of constituents. Additionally, the chains with an odd number of constituents overlap with a radially excited spherical single boson star. We will investigate whether similar results can be obtained for the rotating case. Furthermore, in rotating spacetimes such as Kerr black holes and Kerr black holes with scalar hair \cite{Herdeiro:2014goa, Herdeiro:2014jaa, Wang:2018xhw, Delgado:2019prc, Kunz:2019sgn, Kunz:2019bhm}, ergosphere is known to exist. This paper will examine whether the ergosphere also emerges for chains of rotating mini-boson stars.

The structure of this paper is as follows. In Sec. \ref{sec2}, we will introduce the model for chains of rotating mini-boson stars without self-interacting, including the ansatz for the metric and scalar field, as well as the equations of motion. In Sec. \ref{sec3}, we will present the boundary conditions and outline how we obtain some quantities of interest. In Sec. \ref{sec4}, we will provide numerical results of chains of rotating mini-boson stars with one to five constituents. In the final section, we make a summary and offer prospects for future research.

\section{The model setup}\label{sec2}
We consider Einstein's gravity coupled minimally to a free, complex massive scalar field in a $(3+1)$-dimensions space-time, with the action
\begin{equation}
\label{S}
\mathcal{S} =\int_{\Omega} d^4x \sqrt{-g}\left(\frac{R}{16\pi G}-\nabla_a\psi^*\nabla^a\psi-\mu^2|\psi|^2\right)\,,
\end{equation}
where $R$ is the Ricci scalar curvature, $G$ is Newton's constant and $\mu$ is the mass of scalar field. By varying the action, we can obtain the Klein-Gordon equation
\begin{equation}
\label{KGE}
\Box\psi=\mu^2\psi\,,
\end{equation}
where $\Box$ is the covariant D'Alembert differential operator. And Einstein equations read as
\begin{equation}
\label{EE}
E_{ab}\equiv R_{ab}-\frac{1}{2}g_{ab}R-8 \pi G~T_{ab}=0 \ ,
\end{equation}
where
\begin{equation}
\label{Tab}
T_{ab}\equiv
 \nabla_a\psi^*\nabla_b\psi
+\nabla_b\psi^*\nabla_a\psi
-g_{ab}  (\nabla_c\psi^*\nabla^c\psi+\mu^2|\psi|^2) \, ,
\end{equation}
is the energy-momentum tensor of the scalar field.

To obtain stationary solutions of chains of rotating mini-boson stars, we also adopt the axisymmetric metric with Kerr-like coordinates within the following ansatz:
\begin{eqnarray}
\label{ds2}
ds^2 = -e^{2F_0(r,\theta)}dt^2 + e^{2F_1(r,\theta)}\left( dr^2 + r^2 d\theta^2 \right) + e^{2F_2(r,\theta)}r^2\sin^2\theta \left( d\varphi - W(r,\theta)dt \right)^2 \ .
\end{eqnarray}
In addition, the ansatz of the complex scalar field is given by
\begin{eqnarray}
\label{phi}
\psi=\phi(r,\theta) e^{i(m\varphi-\omega t)}, \;\;\;  m=\pm1,\pm2,  \cdots .
\label{scalar_ansatz}
\end{eqnarray}
Here, the five functions $F_{i}(r,\theta)~(i=0,1,2)$, $W(r,\theta)$ and $\phi(r,\theta)$ depend on the radial distance $r$ and polar angle $\theta$. The constant $\omega$ is the frequency of the complex scalar field and $m$ is the azimuthal harmonic index.

\section{Boundary conditions}\label{sec3}

Before numerically solving these differential equations, we should give some appropriate boundary conditions to describe the asymptotic behaviors of the metric functions $F_{i}(r,\theta)~(i=0,1,2)$, $W(r,\theta)$ and the scalar field $\phi(r,\theta)$.

Because of asymptotic flatness at spatial infinity, the boundary conditions are 
\begin{equation}
\label{br1}
  F_0\bigl.\bigr|_{r\to \infty}=F_1\bigl.\bigr|_{r\to \infty}=F_2\bigl.\bigr|_{r\to \infty}=W\bigl.\bigr|_{r\to \infty}=\phi\bigl.\bigr|_{r\to \infty}=0.
\end{equation}
For axial symmetry, the boundary conditions on the axis ($\theta=0,\pi$) are
\begin{equation}
\label{bth0}
\partial_\theta F_0(r, 0)=\partial_\theta F_1(r, 0)=\partial_\theta F_2(r, 0) =\partial_\theta W(r, 0)=\phi(r, 0)=0.
\end{equation}
At the origin, we require
\begin{equation}
\label{br0}
\partial_r F_0(0,\theta)=\partial_r F_1(0,\theta)=\partial_r F_2(0,\theta) =\partial_r W(0,\theta)=\phi(0,\theta)=0.
\end{equation}
Here, the values of $F_{i}(0,\theta)~(i=0,1,2)$ and $W(0,\theta)$ are the constants independent of the angle $\theta$.

Also, all the solutions discussed in this work have polar angle reflection symmetry $\theta\rightarrow\pi-\theta$ on the equatorial plane. So, we just need to consider the range $\theta \in [0,\pi/2] $ for the angular variable. For different solutions, the scalar field can have either even or odd parity on the equatorial plane. If there is an odd number of boson stars on the chain, the scalar field is even parity, and the boundary conditions are 
\begin{equation}
\label{btheven}
\partial_\theta F_0(r,\pi/2)=\partial_\theta F_1(r,\pi/2)=\partial_\theta F_2(r,\pi/2) = \partial_\theta W(r,\pi/2)=\partial_\theta \phi(r,\pi/2) = 0.
\end{equation}
If there is an even number of boson stars on the chain, the scalar field is odd parity, and the boundary conditions are
\begin{equation}
\label{bthodd}
\partial_\theta F_0(r,\pi/2)=\partial_\theta F_1(r,\pi/2)=\partial_\theta F_2(r,\pi/2) = \partial_\theta W(r,\pi/2)= \phi(r,\pi/2) = 0.
\end{equation}

Near the boundary $r\rightarrow\infty$, we can extract the ADM mass  $M$ and total angular momentum $J$ from the following expanded form of the metric functions $g_{tt}$ and $g_{\varphi t}$
\begin{eqnarray}
\label{ADM}
g_{tt}= -1+\frac{2GM}{r}+\cdots, \nonumber\\
g_{\varphi t}= -\frac{2GJ}{r}\sin^2\theta+ \cdots.
\end{eqnarray}

\section{Numerical results}\label{sec4}

In this section, we will solve the Einstein-Klein-Gordon equations and obtain solutions for chains of rotating mini-boson stars with one to five constituents. The parity of the solutions depends on the choice of boundary conditions (\ref{btheven}) or (\ref{bthodd}). There are four input parameters, corresponding to Newton’s constant $G$, the field mass $\mu$, the frequency $\omega$, and the azimuthal harmonic index $m$. We take the following field redefinition and scaling of both $r$ and $\omega$:
\begin{equation}
\label{sca}
\phi \rightarrow \frac{\phi}{\sqrt{8\pi G}},\quad W \rightarrow \mu W,\quad r \rightarrow \frac{r}{\mu},\quad \omega \rightarrow \mu\omega .
\end{equation}
In all the following results, we set $m=1$. As such, we have only one input parameter $\omega$. For simplicity, we set $G=1$ in these quantities of interest. To facilitate this, we introduce a new coordinate $x$, transforming the radial coordinate range from $[0,\infty)$ to $[0, 1]$,
\begin{equation}
\label{rtx}
x=\frac{r}{r+1}.
\end{equation}
Utilizing reflection symmetry on the equatorial plane ($\theta \rightarrow \pi-\theta$), we can consider the angular variable within the range $\theta \in [0,\pi/2]$. All numerical computations are based on the finite element method, with a grid size of $200 \times 100$ for the integration region $0 \leq x \leq 1$ and $0 \leq \theta \leq \pi/2$. The iteration process employs the Newton-Raphson method, and the relative error estimation for the numerical solutions in this paper is below $10^{-5}$.  For better visualization, we will use coordinates $\rho= r \sin\theta$ and $z= r \cos\theta$ in the subsequent presentation of results.

\begin{figure}[h!]
    \begin{center}
        \includegraphics[height=.22\textheight]{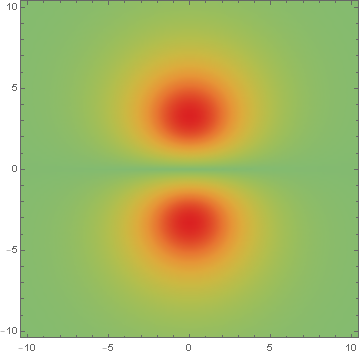}
        \includegraphics[height=.22\textheight]{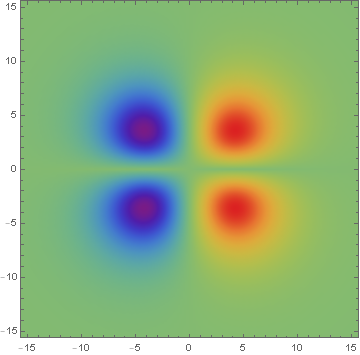}
        \includegraphics[height=.22\textheight]{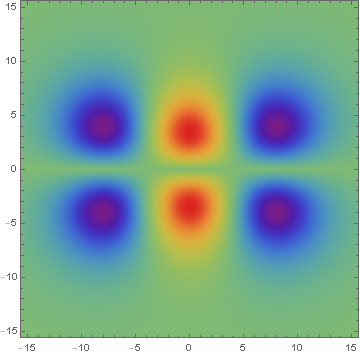}
        \includegraphics[height=.22\textheight]{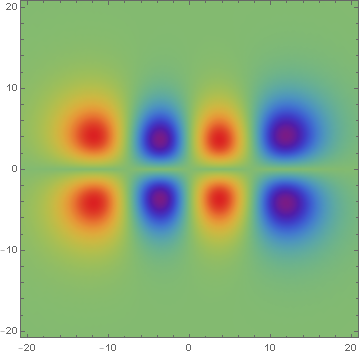}
        \includegraphics[height=.22\textheight]{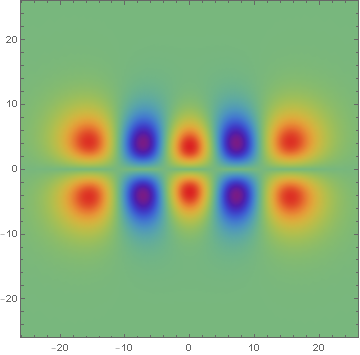}
    \end{center}
    \caption{\small
The scalar field functions $\phi$ for chains of rotating mini-boson stars with one to five constituents in the $z-\rho$ plane (The horizontal axis is the z-axis, and the vertical axis is the $\rho$-axis. We continue to use this setting in the following figures.) for $\omega = 0.85$. The colors close to red have the positive $\phi$, and the colors close to purple have the negative $\phi$.}
\label{fphi}
\end{figure}

\begin{figure}[h!]
    \begin{center}
        \includegraphics[height=.16\textheight]{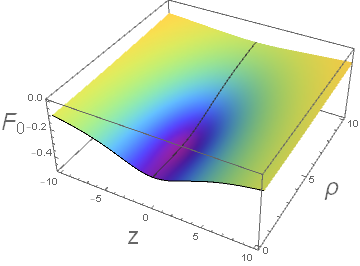}
        \includegraphics[height=.16\textheight]{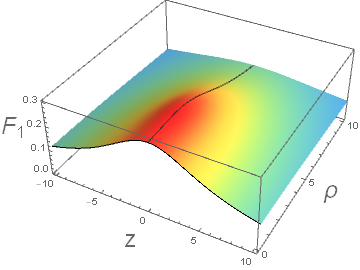}
        \includegraphics[height=.16\textheight]{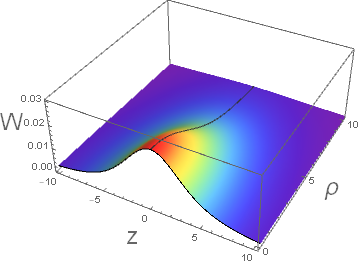}
        \includegraphics[height=.16\textheight]{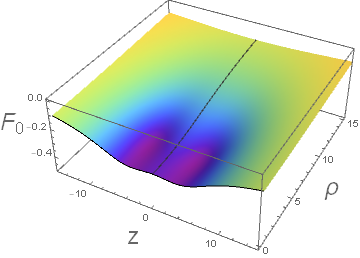}
        \includegraphics[height=.16\textheight]{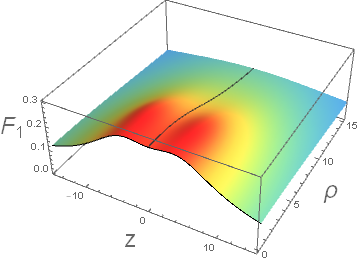}
        \includegraphics[height=.16\textheight]{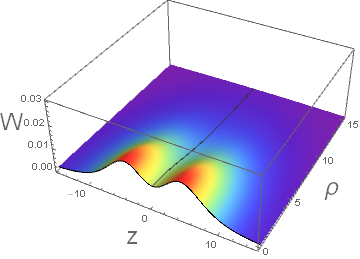}
        \includegraphics[height=.16\textheight]{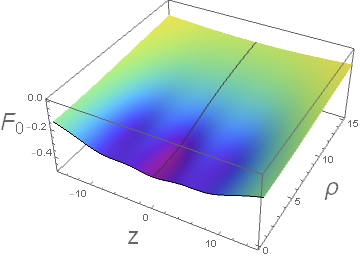}
        \includegraphics[height=.16\textheight]{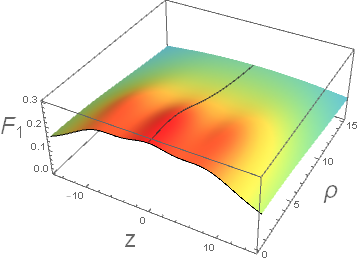}
        \includegraphics[height=.16\textheight]{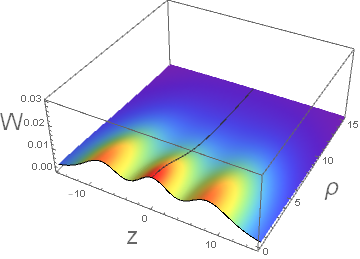}
        \includegraphics[height=.16\textheight]{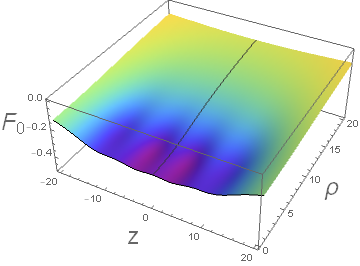}
        \includegraphics[height=.16\textheight]{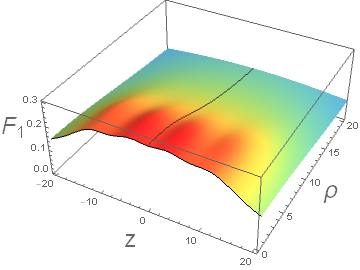}
        \includegraphics[height=.16\textheight]{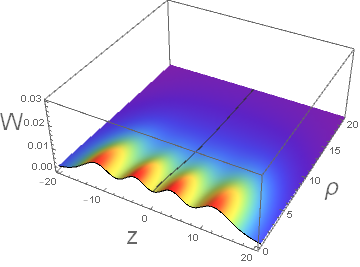}
        \includegraphics[height=.16\textheight]{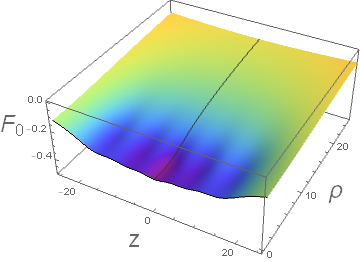}
        \includegraphics[height=.16\textheight]{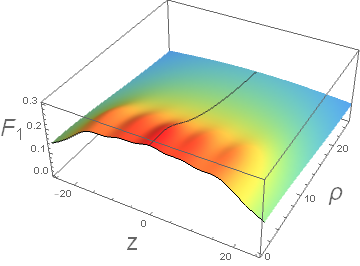}
        \includegraphics[height=.16\textheight]{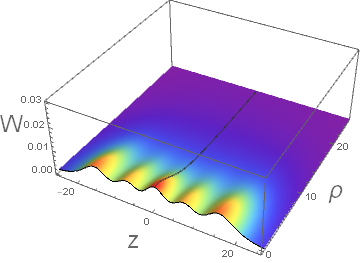}
    \end{center}
    \caption{\small
The metric functions $F_0$ (left), $F_1$ (middle), and $W$ (right) of the chain of rotating mini-boson stars with one to five constituents (from top to bottom) in the $z-\rho$ plane for $\omega = 0.85$.}
\label{fmetr}
\end{figure}

In Fig. \ref{fphi}, we show the distribution of the scalar field in the $z-\rho$ plane. Multiple rotating mini-boson stars are arranged consecutively along their common rotation axis, and all solutions in the figure have a scalar field frequency $\omega =0.85$. As the shape of a single rotating boson star is like a torus, it appears as two circular forms in the $z-\rho$ plane. It can be observed that, for each chain of rotating mini-boson stars, every constituent has nearly size. For chains with an even number of boson stars, the scalar field is antisymmetric about the equatorial plane, while for chains with an odd number of boson stars, it is symmetric. We also show the distribution of metric functions in Fig. \ref{fmetr} corresponding to these solutions in Fig. \ref{fphi}. It can be observed that the metric functions of chains of rotating mini-boson stars are symmetric about the equatorial plane. The minimum value of $F_0$ and the maximum value of $F_1$ do not vary significantly with the number of constituents. However, the maximum value of $W$ decreases with an increasing number of constituents. For chains of rotating mini-boson stars with more than two constituents, the local minima of $F_0$ near the center are smaller than those farther away from the center, as well as the local maxima of $F_1$ near the center are larger than those farther away from the center.

\begin{figure}[h!]
    \begin{center}
        \includegraphics[height=.26\textheight]{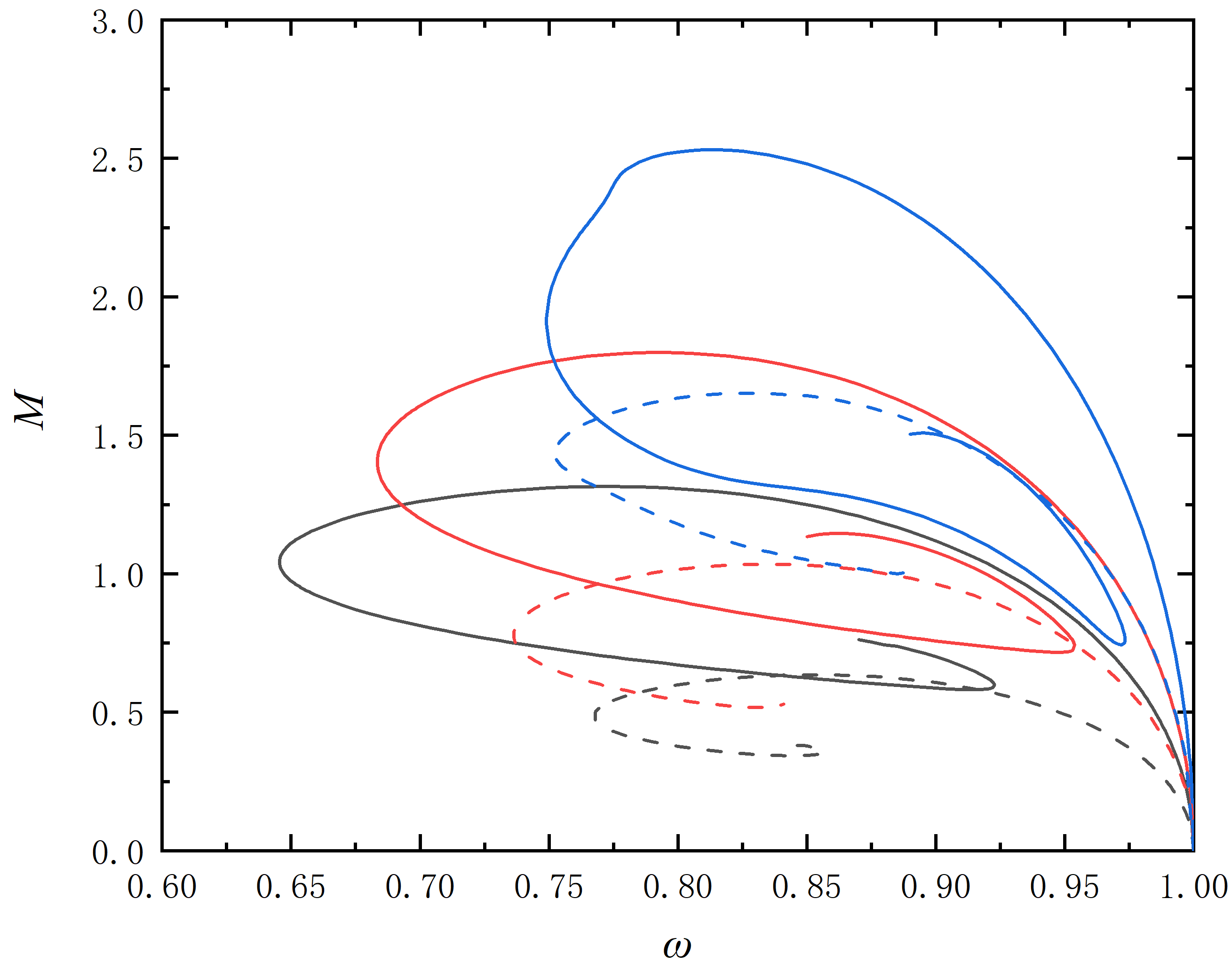}
        \includegraphics[height=.26\textheight]{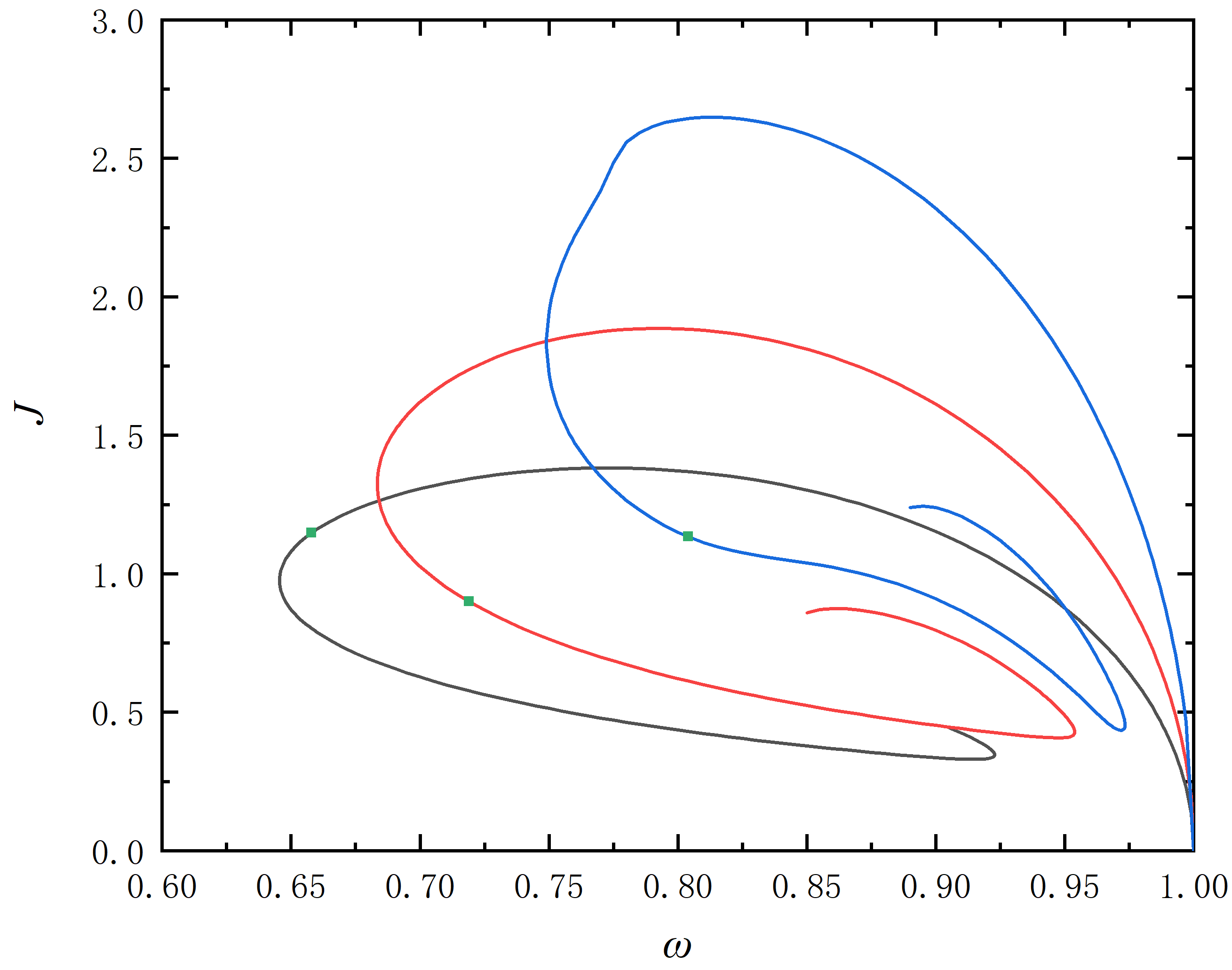}
    \end{center}
    \caption{\textit{Left}: The ADM mass $M$ as a function of the frequency $\omega$. \textit{Right}: The total angular momentum $J$ as a function of the frequency $\omega$. The black, red, and blue lines correspond to chains with one, two, and four constituents, respectively. The solid lines correspond to the rotating case, while the dashed lines correspond to the non-rotating case. The green dots correspond to solutions where ergospheres just appear.}
\label{f124}
\end{figure}

The $\omega$-dependences for chains with one, two, and four constituents are shown in Fig. \ref{f124}. It can be observed that both $M$ and $J$ exhibit the same variation pattern with frequency. At the maximum value of frequency, both ADM mass and total angular momentum vanish, corresponding to the Minkowski vacuum. Starting from the vacuum solution, as the frequency decreases, both $M$ and $J$ increase, reaching their maximum values, and then decrease with further decreasing frequency. We refer to this part of the curve as the first branch. In the second branch, starting from the minimum value of frequency, mass and angular momentum first decrease, then increase, reaching a turning point before entering the third branch. The trend in the third branch is similar to the first. The overall curve forms a spiral shape. It can be observed that, as the number of constituents increases, the curves shift upward. These features are consistent with the non-rotating case. For comparison with the non-rotating case, we show the $M-\omega$ curve of chains of non-rotating mini-boson stars with one, two, and four constituents in the left panel of Fig. \ref{f124}. It can be seen that there are some differences between the rotating case and the non-rotating case. The minimum frequencies of rotating solutions with two and four constituents are larger than that of the single rotating boson star. However, the minimum frequencies of non-rotating solutions with two and four constituents are smaller than that of the single boson star.

\begin{figure}[h!]
    \begin{center}
        \includegraphics[height=.26\textheight]{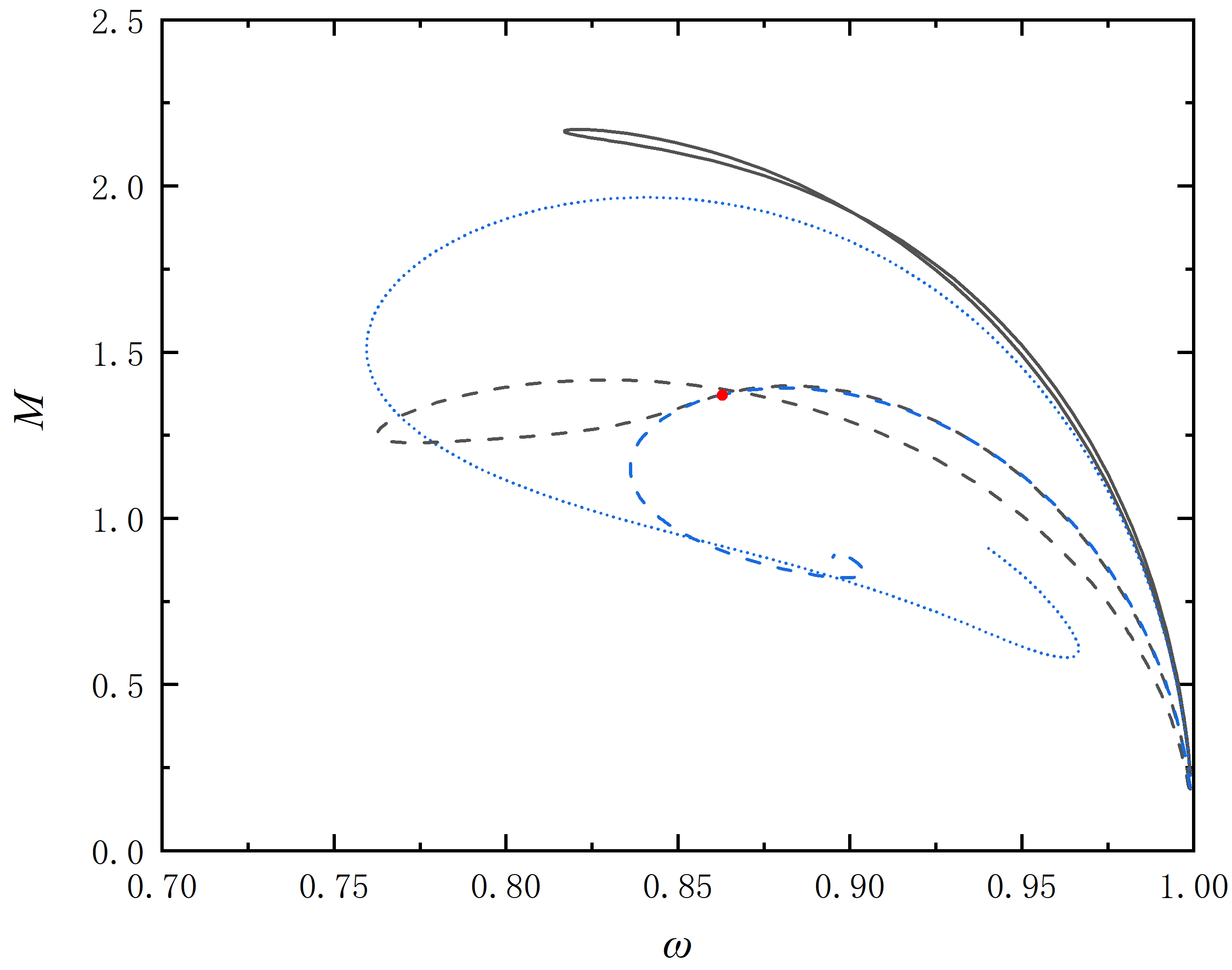}
        \includegraphics[height=.26\textheight]{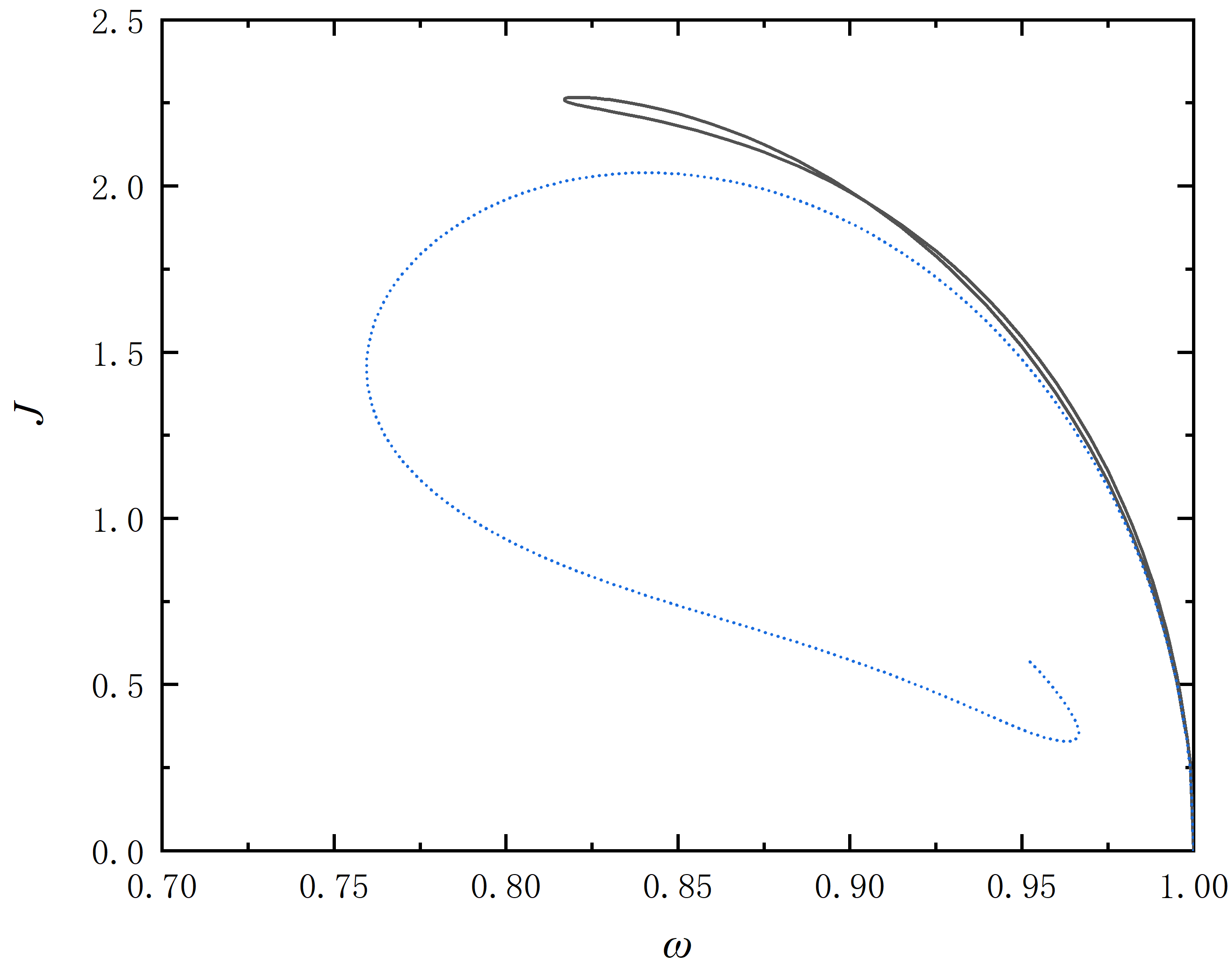}
    \end{center}
    \caption{\textit{Left}: The mass $M$ as a function of the frequency $\omega$. \textit{Right}: The total angular momentum $J$ as a function of the frequency $\omega$. The black solid line corresponds to chains of rotating mini-boson stars with three constituents, and the blue short dot line corresponds to the exited rotating BSs with a radial node. The black dash line corresponds to chains of non-rotating mini-boson stars with three constituents, and the blue dash line corresponds to the exited mon-rotating BSs with a radial node.}
\label{f3ex}
\end{figure}

The $M-\omega$ and $J-\omega$ are displayed in Fig.\ref{f3ex}. For the same motivation as in Fig.\ref{f124}, we also show $M-\omega$ for the non-rotating case. From Fig.\ref{f3ex}, we can see the same feature between the rotating case and the non-rotating case. The ADM mass $M$ and the total angular momentum $J$ of the chains with three constituents have a nontrivial loop structure. The curves start from the vacuum solution, and as the frequency decreases, both $M$ and $J$ increase, reaching their maximum values and then decreasing. In the second branch, both $M$ and $J$ of the rotating case monotonically decrease with increasing frequency eventually returning to the vacuum solution. For the non-rotating case, $M$ will increase to a local maximum before decreasing to the vacuum solution. There also exist other differences. In the non-rotating case, the $M-\omega$ curve of the chains with three constituents overlaps with a first excited spherical single boson star. The red dot is the point where they intersect. For the rotating case, there is no intersection between chains and the first excited state boson star. Furthermore, the minimum frequency of chains with three constituents is larger than that of the first excited state boson star, opposite to the non-rotating case.

\begin{figure}[h!]
    \begin{center}
        \includegraphics[height=.22\textheight]{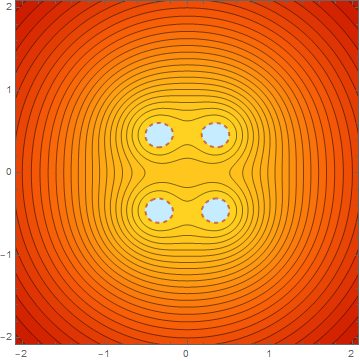}
        \includegraphics[height=.22\textheight]{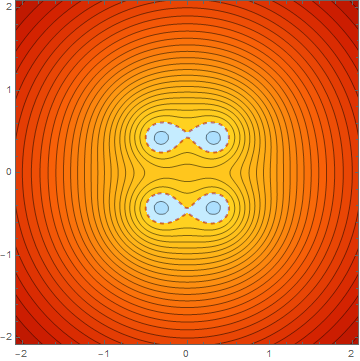}
        \includegraphics[height=.22\textheight]{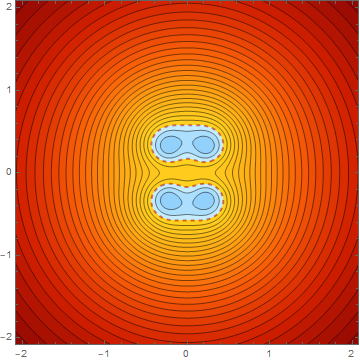}
    \end{center}
    \caption{Isocontours of $g_{tt}$ for chains of rotating mini-boson stars with two constituents in the $z-\rho$ plane. The three panels from left to right correspond to solutions with $\omega=0.74$, $\omega=0.76$ and $\omega=0.85$ in the second branch. Warm colors represent negative $g_{tt}$, cold colors represent positive $g_{tt}$, and the red dash line represents ergosurface ($g_{tt}=0$).}
\label{f2}
\end{figure}

\begin{figure}[h!]
    \begin{center}
        \includegraphics[height=.22\textheight]{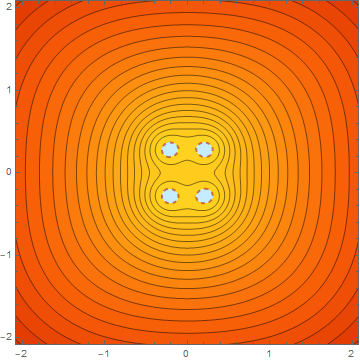}
        \includegraphics[height=.22\textheight]{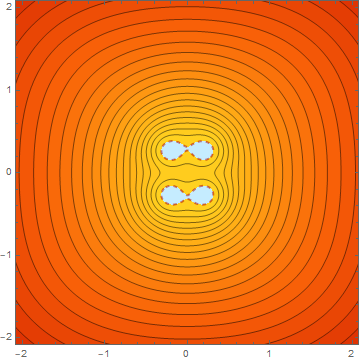}
        \includegraphics[height=.22\textheight]{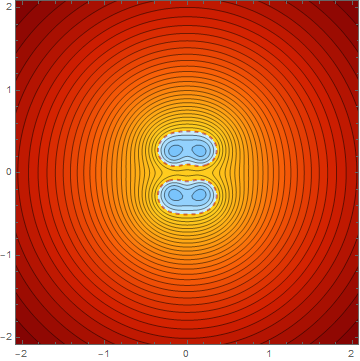}
    \end{center}
    \caption{Isocontours of $g_{tt}$ for chains of rotating mini-boson stars with four constituents in the $z-\rho$ plane. The three panels from left to right correspond to solutions with $\omega=0.815$, $\omega=0.825$, and $\omega=0.950$ in the second branch. Warm colors represent negative $g_{tt}$, cold colors represent positive $g_{tt}$, and the red dash line represents ergosurface ($g_{tt}=0$).}
\label{f4}
\end{figure}

In rotating spacetime, the ergosphere may appear. We find there exist ergosphere for some chains of rotating mini-boson stars with one, two, and four constituts. In the right panel of Fig. \ref{f124}, we use the green dots to mark these solutions where the ergospheres just appear. For the single rotating mini-boson stars, an ergosphere appears as the frequency decreases to $\omega=0.658$ in the first branch. For chains of rotating mini-boson stars with two and four constituents, two ergospheres appear as the frequency increased to $\omega=0.719$ and $\omega=0.804$ in the second branch, respectively. With increasing frequency, these two regions merged into a single ergosphere. To illustrate this change clearly, we presented the isocontours of $g_{tt}$ in the $z-\rho$ plane for chains of rotating mini-boson stars with two and four constituents in Fig. \ref{f2} and Fig. \ref{f4}, respectively. Warm colors represent negative $g_{tt}$, while cool colors represent positive $g_{tt}$ (i.e., the ergosphere). For chains of rotating mini-boson stars with three and five constituents, there is no ergosphere.

\begin{figure}[h!]
    \begin{center}
        \includegraphics[height=.21\textheight]{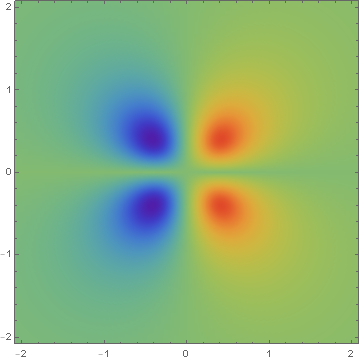}
        \includegraphics[height=.21\textheight]{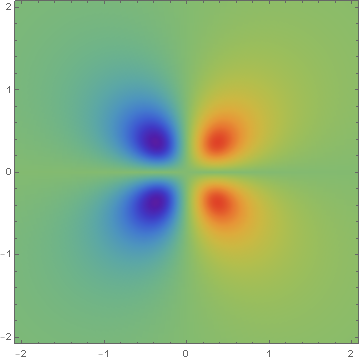}
        \includegraphics[height=.21\textheight]{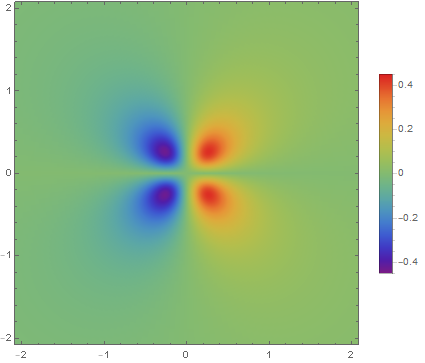}
    \end{center}
    \caption{Distributions of the scalar field function $\phi$ for chains of rotating mini-boson stars with two constituents in the $z-\rho$ plane. The three panels from left to right correspond to solutions with $\omega=0.74$, $\omega=0.76$, and $\omega=0.85$ in the second branch.}
\label{f2p}
\end{figure}

\begin{figure}[h!]
    \begin{center}
        \includegraphics[height=.21\textheight]{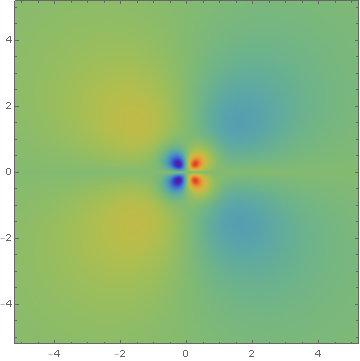}
        \includegraphics[height=.21\textheight]{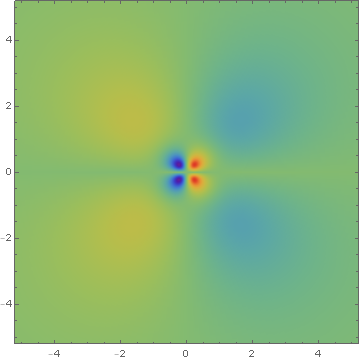}
        \includegraphics[height=.21\textheight]{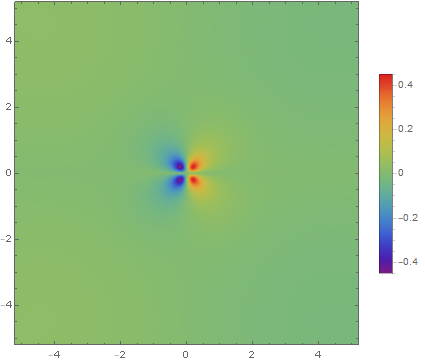}
    \end{center}
    \caption{Distributions of the scalar field function $\phi$ for chains of rotating mini-boson stars with four constituents in the $z-\rho$ plane. The three panels from left to right correspond to solutions with $\omega=0.815$, $\omega=0.825$, and $\omega=0.950$ in the second branch.}
\label{f4p}
\end{figure}

To explore the reasons behind these variations in the ergosphere, we show distributions of scalar field function $\phi$ in Fig. \ref{f2p} and Fig. \ref{f4p}, with the same solutions from Fig. \ref{f2} and Fig. \ref{f4}. From Fig. \ref{f2p}, we can see that the absolute value of extrema of scalar field function increases with the frequency, and the distance between the positions of the two extrema decreases as the frequency increases. This indicates that the scalar field concentrates toward the center as the frequency increases. It can be seen in Fig. \ref{f4p}, for chains of rotating mini-boson stars with four constituent, the absolute value of the extrema away from the center is much smaller than that near the center. This elucidates why, for chains with four constituents, there are only two ergospheres instead of four. Additionally, the variation of the extrema near the center with frequency is the same as that of chains with two constituents. This also portends that the two ergospheres will expand and approach each other, eventually merging into a single ergosphere.

\section{Conclusion}\label{sec5}

In this article, we extend chains of mini-boson stars to the rotating case. These are soliton solutions formed by the coupling of non-interacting scalar fields with Einstein gravity. This implies that in rotating spacetime, solutions with chain configurations can be obtained even in the absence of self-interaction, because of the presence of Einstein gravity.

We obtained solutions of the chains of rotating mini-boson stars with one to five constituents numerically and presented the distribution of metric field functions and scalar field functions for these solutions. For chains with an odd number of constituents, the scalar field functions are symmetric about the equatorial plane, while for chains with an even number of constituents, they are anti-symmetric about the equatorial plane. The metric field functions are all symmetric about the equatorial plane. We also obtained the relations between the ADM mass $M$, total angular momentum $J$, and frequency $\omega$ for these solutions. For chains of rotating mini-boson stars with one, two, and four constituents, the curves exhibit a spiral shape, while for chains with three constituents, the curves form a loop. This is consistent with the non-rotating case. There are some differences between the rotating case and the non-rotating case. For example, there is no intersection between chains with an odd number of constituents and the excited state boson star with radial nodes. 

We also analyzed the ergosphere for chains of rotating mini-boson stars. For chains with two and four constituents, as the frequency changes, two ergospheres appear and then merge into a single ergosphere. We also show distributions of the scalar function $\phi$ of solutions that have ergosphere. For single boson stars, solutions closer to the center of the spiral $M-\omega$ curve have higher compactness. Therefore, we conjecture that for chains with an even number of constituents which have spiral $M-\omega$ curves, ergospheres emerge. And because of the scalar field is odd-parity, there exist two ergospheres when ergospheres just appear. For chains with an odd number of constituents which have loop $M-\omega$ curves, there is no ergosphere, because these solutions are not compact enough.

This work can have some interesting extensions. Firstly, since these chain solutions can also be considered as excited state solutions, it is possible to obtain multistate solutions where different scalar fields are in different excited states. Secondly, considering geodesics in the background of rotating boson star chains may yield some intriguing orbits. Finally, reference \cite{Herdeiro:2023roz} obtained stationary solutions of two Kerr black holes with scalar hair through rotating dipolar boson stars (i.e., chains of rotating boson stars with two constituents). It is possible that obtaining stationary solutions of multi Kerr black holes through chains of rotating boson stars with more constituents.

\section*{ACKNOWLEDGEMENTS}\label{ack}

This work is supported by the National Key Research and Development Program of China (Grant No. 2020YFC2201503) and the National Natural Science Foundation of China (Grants No. 12275110 and No. 12247101).

\providecommand{\href}[2]{#2}\begingroup\raggedright
\endgroup

\end{document}